\documentclass[9pts,conference,hidelinks]{IEEEtran}
\IEEEoverridecommandlockouts
\usepackage{cite}
\usepackage{amsmath,amssymb,amsfonts}
\usepackage{graphicx}
\usepackage{textcomp}
\usepackage{balance}
\usepackage{xcolor}
\usepackage{algorithm}
\usepackage[noend]{algpseudocode}
\usepackage{mathtools}
\usepackage{multirow}
\usepackage{enumitem}
\usepackage{caption}
\usepackage{makecell}
\usepackage{subcaption}
\usepackage{calc}
\usepackage{comment}
\usepackage{etoolbox}
\usepackage{booktabs}
\usepackage{adjustbox}
\usepackage{listings}
\usepackage{xcolor}
\usepackage[colorinlistoftodos]{todonotes}

\usepackage{url}

\usepackage{breakurl}
\usepackage[breaklinks]{hyperref}
\hypersetup{colorlinks,
}
\usepackage{siunitx}

\definecolor{codegreen}{rgb}{0,0.6,0}
\definecolor{codegray}{rgb}{0.5,0.5,0.5}
\definecolor{codepurple}{rgb}{0.58,0,0.82}
\definecolor{backcolour}{rgb}{0.95,0.95,0.92}

\lstdefinestyle{mystyle}{
    backgroundcolor=\color{backcolour},   
    commentstyle=\color{codegreen},
    keywordstyle=\color{magenta},
    numberstyle=\tiny\color{codegray},
    stringstyle=\color{codepurple},
    basicstyle=\ttfamily\footnotesize,
    breakatwhitespace=false,         
    breaklines=true,                 
    captionpos=b,                    
    keepspaces=true,                 
    numbers=left,                    
    numbersep=5pt,                  
    showspaces=false,                
    showstringspaces=false,
    showtabs=false,                  
    tabsize=2,
    moredelim=**[is][\color{red}]{@}{@}1
}
\def\BibTeX{{\rm B\kern-.05em{\sc i\kern-.025em b}\kern-.08em
    T\kern-.1667em\lower.7ex\hbox{E}\kern-.125emX}}
\begin{document}

\title{\LARGE 
Challenges and Opportunities to Enable Large-Scale Computing  
\\ via Heterogeneous Chiplets
\\
\large Invited Paper

\vspace{-20pt}
}
\vspace{-16pt}
\author{Zhuoping Yang\IEEEauthorrefmark{1}, Shixin Ji\IEEEauthorrefmark{1}, Xingzhen Chen\IEEEauthorrefmark{1}, Jinming Zhuang\IEEEauthorrefmark{1}, Weifeng Zhang\IEEEauthorrefmark{2}, Dharmesh Jani\IEEEauthorrefmark{3}, Peipei Zhou\IEEEauthorrefmark{1}  \\
\IEEEauthorrefmark{1} University of Pittsburgh,
\IEEEauthorrefmark{2} Lightelligence,
\IEEEauthorrefmark{3} Meta\\
Email: {\tt \{zhuoping.yang, peipei.zhou\}@pitt.edu}\IEEEauthorrefmark{1}, \\ {\tt weifeng.zhang@lightelligence.ai}\IEEEauthorrefmark{2} ,  {\tt janidb@meta.com}\IEEEauthorrefmark{3}
}

\makeatletter
\patchcmd{\@maketitle}
  {\addvspace{0.5\baselineskip}\egroup}
  {\addvspace{-1.0\baselineskip}\egroup}
  {}
  {}
\makeatother

\maketitle
\setlength{\textfloatsep}{0pt}

\begin{abstract}

Fast-evolving artificial intelligence (AI) algorithms such as large language models have been driving the ever-increasing computing demands in today's data centers. 
Heterogeneous computing with domain-specific architectures (DSAs) brings many opportunities when scaling up and scaling out the computing system. 
In particular, heterogeneous chiplet architecture is favored to keep scaling up and scaling out the system as well as to reduce the design complexity and the cost stemming from the traditional monolithic chip design. 
However, how to interconnect computing resources and orchestrate heterogeneous chiplets is the key to success. 
In this paper, we first discuss the diversity and evolving demands of different AI workloads. 
We discuss how chiplet brings better cost efficiency and shorter time to market.
Then we discuss the challenges
in establishing chiplet interface standards, packaging, and security issues.
We further discuss the software programming challenges in chiplet systems.
\end{abstract}


\section{Introduction}
\label{sec:intro}


Artificial intelligence (AI) and deep learning (DL) have provided an effective way to address complicated tasks in applications including computer vision, natural language processing, etc. 
Many hardware accelerators including GPUs, dedicated DL application-specific integrated circuits (ASICs), and FPGAs, are proposed to increase both throughput and energy efficiency. 
GPUs achieve high throughput by massive parallelism, and they are widely used in deep neural network (DNN) training as they can hugely exploit the parallelism in large batches of training data. 
From the deep learning inference aspect, for example in real-time application scenarios where low latency is required, ASICs are designed in the seek for better customization. 
However, designing ASICs is not trivial. 
Companies would spend three or four years developing their first ASIC from scratch, and two or three years releasing a subsequent ASIC~\cite{Reuther2022trends}, not to mention the huge costs in dollars (millions to billions).
FPGAs offer greater programmability compared to ASICs, enabling rapid and economical product updates, albeit with some trade-offs in performance. 
In fact, to achieve higher energy efficiency for DL tasks while providing flexibility, there is a trend that CPUs, FPGAs, GPUs, and accelerators (Accs) are integrated into the same system-on-chip (SoC). 
Intel FPGAs Stratix 10 NX FPGA~\cite{intel_s10_fpga} and AMD Versal ACAP architecture~\cite{fpga19versal,fpga23charm,iccad23aim,dac23automm} integrate AI tensor cores into the FPGA fabric (FPGA+Acc). 
Intel CPUs integrate vector units to support SSE or AVX instructions. 
The latest AMD  Ryzen™ AI CPUs also integrate dedicated AI engines in the CPUs (CPU+Acc)~\cite{amd_ryzen_ai}. 
Nvidia Jetson Orin GPUs have integrated tensor cores, multimedia cores, and NVDLA with the CUDA cores (GPU+CPU+Accs).

In the abovementioned heterogeneous systems, integration can be achieved with intellectual properties (IPs) designed by the same company or through licensed IPs from other companies. 
The emerging chiplet technologies are enabling novel heterogeneous integration across different IP vendors. 
Traditional chip is implemented on a monolithic silicon die but the die size is approaching the lithographic reticle limits due to growing complicated functionality and slow down of process technology. 
Chiplets are smaller chips disaggregated from an SoC and optimized for in-package communication~\cite{hutner2020special}. 
In the vision of the Open Compute Project subgroup Open Domain Specific Architecture (ODSA)~\cite{odsa}, chiplets can be easily reused and integrated on an interposer even if chiplets are manufactured by different vendors. 
However, there remain enormous research questions to be explored. 
In this paper, we focus on how chiplets can meet new demands in generative AI workload, and discuss the challenges in both hardware and software 
with potential solutions. 

\section{Infrastructure Challenges from Diverse and Evolving AI Workloads}
\label{sec:challenge}


\setlength{\textfloatsep}{0pt}
\begin{table*}[!tbh]
\footnotesize
\caption{Comparisons between chiplet and PCB, monolithic ASIC~\cite{li2020chiplet}.}
\vspace{-5pt}
\label{tbl:tech_comparison}
\begin{adjustbox}{width=2\columnwidth,center}
\begin{tabular}{c | c c c c c}
 \toprule

   \textbf{Integration Technology}  
  &  \textbf{Design Cycle} &  \textbf{Cost/\$}  &  \textbf{Integration} &  \textbf{Energy Efficiency} &  \textbf{Performance}\\
    \midrule
   
   Monolithic ASIC    & \textgreater 1 year   & \textgreater 1,000,000 & +++ & +++ & +++  \\
   Chiplet   & months   & 1,000-1,000,000 & ++ & ++ & ++  \\
   PCB    & weeks  & 100-10,000 & + & + & +  \\
   
    \bottomrule
\end{tabular}
\end{adjustbox}
\vspace{-20pt}
\end{table*}

\begin{figure}[h]
\centering
\includegraphics[width=1\columnwidth]{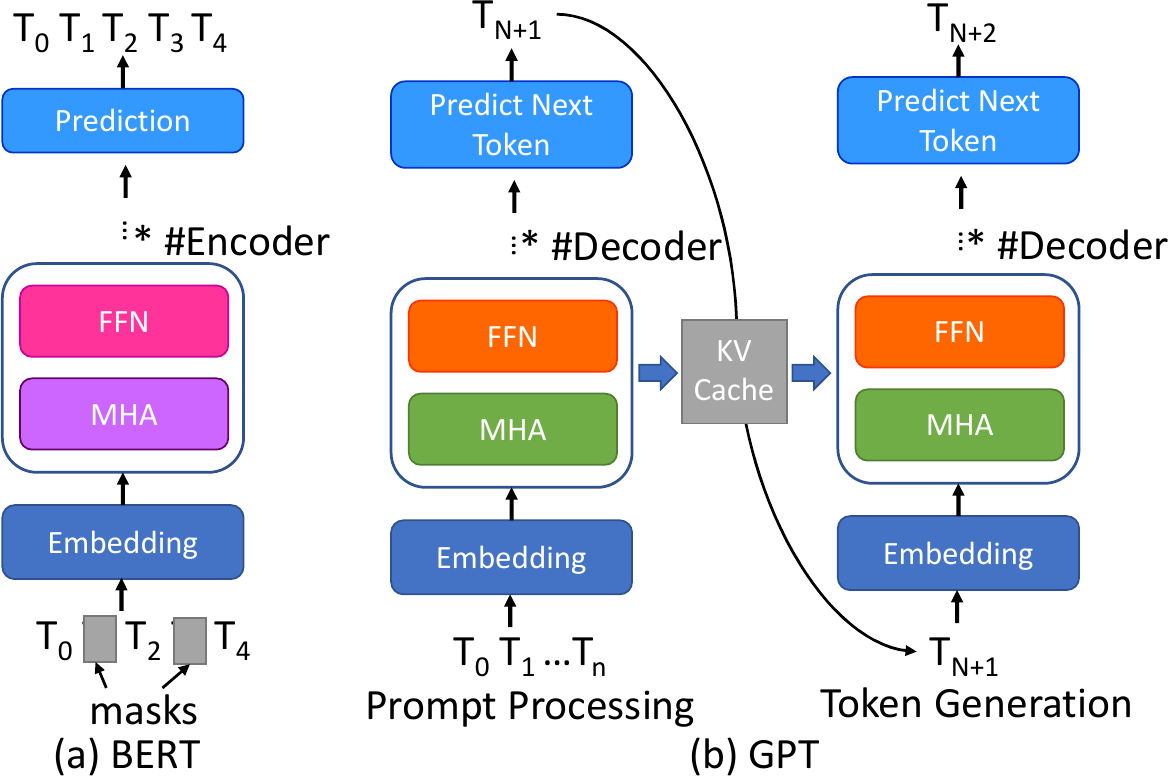}
\vspace{-5pt}
\caption{Different processes in BERT and GPT.}
\label{fig:bert-gpt}
\vspace{-10pt}
\end{figure}

In the acceleration of AI tasks, both communication and computation play crucial roles. 
One way to improve the computation utilization is to overlap the time of communication with the computation.
If the communication time is less than or equal to the computation, no hardware resources are idling. 
However, the communication and computation demands vary in different applications, e.g., GPT\cite{gpt} and BERT\cite{bert}. 
We use arithmetic intensity~\cite{full_stack_transformer1} to characterize the algorithms' data reuse ratio and computation-to-communication (CTC) ratio~\cite{zhang2015optimizing} to characterize the execution time between computation and communication.
\begin{equation}
\begin{gathered}
    Arithmetic~Intensity = \frac{\#Operations}{\#DataMovement} \\
    CTC = \frac{Time_{Comp.}}{Time_{Comm.}} = \frac{\#Operations/HW_{ops}}{\#DataMovement/HW_{bw}}
\end{gathered}
\end{equation}
GPT and BERT are both Transformer model variances. 
As shown in Figure~\ref{fig:bert-gpt} BERT repeats encoder blocks while GPT is built using decoder blocks. 
In generation tasks, the input sequence is first transformed into tokens. 
We can mask some tokens and pass the masked tokens to BERT and BERT predicts all masked tokens in only one inference. 
Different from BERT generation, GPT only generates one next token based on previous tokens and the generation process can be further split into two sub-tasks, prompt processing and token generation. 
In prompt processing, GPT processes all tokens, predicts the next one, and stores intermediate data to avoid recomputation in the token generation processes. 
This technique is referred to as KV-cache. Then, in the token generation, GPT only processes the last token and generates the next token using KV-cache. 
Therefore, GPT has a very low arithmetic intensity in the token generation process 
whereas BERT has a higher arithmetic intensity. For example, when the sequence length is 512, the arithmetic intensity of GPT-2 and BERT-Large are 2 and 266 respectively~\cite{full_stack_transformer1}. 
Similarly, the variances in arithmetic intensity also exist in different operations within the same model. 
Layers including convolution and matrix-multiply are more computationally intensive while activations such as Softmax are more demanding in data movement.
The variances of arithmetic intensity in different workloads entail different demands in bandwidth and computation resources when designing hardware accelerators. 

\vspace{-12pt}
\setlength{\textfloatsep}{0pt}

Additionally, the differences between the growth of peak hardware throughput and bandwidth are growing, which means the design complexity also increases as more data reuse and higher arithmetic intensity are needed.
There is a trend showing that every two years, the peak hardware throughput increases 3x but the bandwidth only increases 1.6x~\cite{gholami2020ai_and_memory_wall}. 
Another trend is that the gap between the hardware memory size and AI model size surges.
In recent two years, the AI model has increased 410x while hardware memory only increases 2x~\cite{gholami2020ai_and_memory_wall}. 
These emerging demands
are pushing innovations in designing scalable hardware acceleration systems.

\section{Chiplets: A Solution for Rapid Heterogeneous System Development}
\label{sec:rapid}

To catch up with the increasing demands, more circuits are integrated into chips. 
However, simply scaling the monolithic ASIC accelerators is hardly practical. 
The chiplet technique is becoming a promising solution to improve performance and energy efficiency and decrease the cost and time to market (design cycle column in Table~\ref{tbl:tech_comparison}), which also suits the different requirements of customers. 

There are two limitations in scaling up monolithic ASIC chips: the chip area and the yield rate.
The die size of CPUs and GPUs grew 14\% and 8\% respectively every year~\cite{DieSize} from 2006 to 2020, which is quickly approaching the lithographic reticle limit.
Furthermore, the yield will decrease with the increase in die sizes, which causes more waste, leading to higher costs. 
One way to keep scaling while minimizing the cost is to fabricate small chips and then integrate them. 
Traditionally, multiple chips can be connected via printed circuit boards (PCBs). 
However, PCB systems are difficult to provide high-density inter-chip connections, which is required in many applications. 
The PCB systems cannot pack more than 400 connections in 1 $cm^2$ because the PCBs are prone to warpage and the distance between solder bumps (used to connect the PCBs and chips) cannot be less than 0.5 mm~\cite{gupta2019goodbye}. 
Besides, due to the wire length and parasitic parameters the PCB systems need more power in interconnection and the signal frequency is low~\cite{hutner2020special}.
The chiplet technique is proposed to keep the performance scaling and maintain a reasonable cost. 
This is done by partitioning a monolithic chip into several smaller chiplets and reassembling them into a system-in-package (SiP). 
This method enables further scaling in the silicon area and decreases the cost. 
Table~\ref{tbl:tech_comparison} compares chiplet technology, monolithic ASIC, and PCB in various metrics~\cite{li2020chiplet}. 


The chiplet technology has been adopted and shown great improvement in real-world products. For example, the first-generation AMD EPYC CPU processor~\cite{naffziger2021pioneering} is based on chiplet architecture, which consists of four chiplets in the 14nm process node. 
Compared with the monolithic design, the chiplet-based architecture integrates more silicon area in total, which exceeds the reticle limits.
Besides, the cost of the chiplet-based architecture is 41\% cheaper than a monolithic design~\cite{DieSize}.

Additionally, chiplet technology allows the integration of chiplets in different process nodes, which provides more choices for different IPs. 
For example, unlike digital circuits, analog circuits or mixed-signal IPs do not benefit greatly from the advanced technology node, and it is also difficult and time-consuming to apply new technology to them~\cite{Furthering}. Therefore, it is necessary to manufacture these IPs in a more mature technology node.
Chiplet allows us to integrate these analog or mixed-signal chiplets together with digital circuit chiplets that are fabricated in more advanced technology.
Besides, designers can select various chiplets to fulfill specific requirements. For example, GPT-based applications have high demands in memory bandwidth~\cite{peng2023chipletcloud}. Therefore, to decrease inference latency, chiplets with more memory channels or higher I/O bandwidth are preferred to be integrated into the SiP.

\begin{table*}[!tb]
\footnotesize
\caption{Comparisons between different chiplet interfaces (data accessed in 2023/11).}
\vspace{-5pt}
\label{tbl:interconnect_comparison}
\begin{adjustbox}{width=2\columnwidth,center}
\begin{tabular}{c | c c c c c}
 \toprule

   \textbf{Protocol}  
  &  \textbf{Institution} &  \textbf{\makecell{Typical Energy \\Efficiency (pJ/bit)}}  &  \textbf{\makecell{Maximum Speed \\ (Gbps/wire)}} &  \textbf{\makecell{Fault Tolerance\\ Mechanism}}\\
    \midrule
   USR~\cite{USR}    & \textbackslash  & $<$0.6 ~\cite{USR}& \textgreater 20~\cite{USR}  & N/A  \\
   AIB~\cite{AiB,AiB1}    & Intel   & 0.85 (Gen1) & 2 (Gen1)~\cite{AiB}, 6.4 (Gen2)~\cite{AiB1} & N/A  \\
   BoW~\cite{BoW,BoW1}   & ODSA   & \textless 0.25-1.0 ~\cite{BoW1} & 32~\cite{BoW}& N/A  \\
   HBM~\cite{HBM}   & JESDC   & \textbackslash & 6.4  & ECC  \\
   LINPINCON~\cite{7936211}   & TSMC   & 0.424 & 2.8 & N/A  \\
   UCIe~\cite{UCIeLink}   & UCIe Union   & 0.25-1.25~\cite{UCIeLink} & 32 GT/s~\cite{UCIeLink} & \makecell{CRC + Retransmission}  \\ 
   AAC~\cite{ACCLink,ACC10}   & \makecell{China Chiplet Industry Alliance}   & 2.5~\cite{ACCLink} & 128~\cite{ACC10}  & \makecell{CRC + BER + Retransmission }  \\
   
    \bottomrule
\end{tabular}
\end{adjustbox}
\vspace{-10pt}
\end{table*}

\section{Hardware Design Challenges}
\subsection {Chiplet Interfaces} 


To achieve heterogeneous integration, further efforts are required for chiplet interconnection design, which includes interconnection protocols standardization, routing algorithms in SiP, and system simulation supports for chiplet system design.

\subsubsection{Chiplet-based protocols and interfaces}

To enable communication between chiplets, chiplets need to follow the same transmission protocol. 
However, different vendors adopt different protocols, which hinders the heterogeneous integration. 
Besides, it is difficult to develop a unified protocol that suits various applications as different applications may have different requirements.

Serial interface and parallel interface are two categories of inter chiplet communication interfaces. 
The serial interface only requires a pair of differential connections to facilitate data transmission in the physical layer, while the parallel interface uses multiple connections~\cite{ardalan2020BoW, ma2022survey}.
For example, ultra-short-reach (USR)~\cite{USR,8741259} is a serial interface~\cite{ma2022survey} designed for die-to-die interconnect, which aims to ultra-short electrical interconnect and has a low power consumption ($<0.6pJ/b$)~\cite{USR}.
Compared with serial interface, parallel interface typically has hundreds of connections and thus can achieve the same bandwidth as the serial interface with a much lower connection rate~\cite{ma2022survey}. 
At present, there are many parallel interface standards, such as Advanced Interface Bus (AIB)~\cite{AiB}, Bunch of Wires (BoW)~\cite{BoW}, high bandwidth memory (HBM) interconnect~\cite{HBM}, Low-voltage-In-Package-INter-CONnect (LINPINCON)~\cite{7936211}, Universal Chiplet Interconnect Express (UCIe)~\cite{UCIe}, Advanced Cost-driven Chiplet (ACC)~\cite{ACC10}. Table~\ref{tbl:interconnect_comparison} shows the differences between the existing chiplet interfaces.
These interfaces provide different bandwidth, latency, etc, and have different demands on the package. 
It is difficult to develop a unified interface that is the best solution for all applications. 
Heterogeneous integration has another opportunity when integrating off-the-shelf chiplets equipped with different chiplet interfaces. 
For example, designing a hub chiplet that communicates to two different off-the-shelf chiplets that are equipped with two different die-to-die (D2D) IPs. However, designers have to pay two times D2D IP non-recurring engineering (NRE) cost for the hub chiplet, which brings extra cost and design complexity.



\subsubsection{Passive and active interposer}
Silicon interposer is one of the chiplet packaging technology.
The passive interposers only have wires while the active interposes have active components and allow offloading digital logic circuits to the interposer. 
The active interposer has many advantages. 
For example, the routers do not have to be placed on the chiplets, which decreases the silicon area of the chiplets. 
Repeaters can be integrated for long signals to improve the frequency.
The active interposer also provides a way to distribute low-jitter and low-skew clocks for all routers~\cite{stow2019investigation}. 
However, there are also design challenges related to the active interposer. 
First, the network topology may have an impact on the performance due to data contention. \cite{bharadwaj2020kite} proposes a set of chiplet interconnect topologies to balance the data traffic, avoid hot spots, and provide better performance. 
Second, the multi-chiplet systems are prone to have deadlock issues, even if each chiplet has been verified for the functionality separately~\cite{taheri2022deft, yin2018modular}. 
The deadlock typically results from the circular data dependencies among different chiplets. 

\subsubsection{Pre-silicon hardware simulator for chiplet-based architecture}

The scale of chiplet-based architecture is much larger than the scale of the monolithic chip. 
Therefore, we need a more powerful, functional, and multi-chiplet scenario-oriented simulator in the pre-silicon phase. 

Though we have several multi-core simulators, we can't use these multi-core simulators for the design of multi-chiplet system~\cite{DesignChallenge}.
In the design of the chiplet-based system, we also need to accurately model the routing layer between several chiplets, which is also not considered in the multi-core simulation. 
There are some existing works on multi-chiplet simulations~\cite{multi-chiplet_methodology, bharadwaj2020kite, krishnan2021siam}. 
As the scale of the chiplet-based systems keeps increasing
we call for more efficient and comprehensive simulators to provide robust support for chiplet-based design.

%

\subsection {Package Related Issues} 

\noindent\textbf{Testing.} 
The chiplet should not be a black box to the system designer. 
The marketplace also needs standards to describe the chiplets in terms of testing, thermal, I/O, etc~\cite{Mastroianni2021standard}. 
Chiplets potentially enable a marketplace where product developers can buy modules for multiple vendors and assemble them at a much lower cost than designing a chip from scratch. 
However, product developers still have to design their own interposer and package. 
A report shows that packaging costs can be comparable with designing a chip due to the low yield of the complex packaging processing and the bounding defects~\cite{Chiplet_actuary}. 
In current packaging technologies, we cannot detach a bonded die from the substrate without damaging the die~\cite{hutner2020special}. 
Therefore, comprehensive tests before the chiplets are assembled are desired. 
These tests on individual chiplets can be provided by the product developers to the packaging vendors and should cover as many user cases as possible to increase the test coverage.
If the testing is performed after the packaging process, and the exact chips with errors are identified, we can not do repackaging as the repackaging process will damage all the good chiplets as well as the package.
However, this is still an open area for academia, as well as industry~\cite{IEEE1687, IEEE1828, Mastroianni2021standard}, and more works to develop the standards which include reducing the required number of pins for testing and standardizing the test interfaces are desired~\cite{hutner2020special}. 

\noindent\textbf{Thermal.} Recent research has shown that the package design cannot guarantee full functionality without considering the chiplets' thermal characteristics~\cite{TAP-2.5D,Kabir2020Holistic, Kabir2020Coupling, Eris2018thermal}. 
Dark silicon describes a phenomenon that in a many-core system, not all cores can be active at the same time due to certain thermal constraints. 
This problem is alleviated in the chiplet-based system. However, we still have to address the thermal issues since chiplets are placed close to each other to minimize the wire length and package size. 
The thermal issues can be addressed using both online management and offline design optimization. 
In run-time management, the computational-intensive tasks can be scheduled to non-adjacent chiplets.
Besides, the thermal issue can be addressed using dynamic voltage and frequency scaling (DVFS) or run-time scheduling after the chiplet systems are manufactured~\cite{Li2022PowerManagement}, but these methods sacrifice performance. 
In offline design optimization, thermal-aware chiplet placement approaches avoid hot spots by increasing distance or inserting low-power chiplets between high-power chiplets~\cite{TAP-2.5D, Eris2018thermal}.
~\cite{TAP-2.5D} shows it can achieve 20
\% peak temperature decrease without sacrificing performance or using expensive cooling technology. 

\noindent\textbf{Package and chiplet co-design.} 
The achievable system performance is determined by both the package design and the chiplet design.
~\cite{Kabir2020Holistic, Kabir2020Coupling} propose a holistic package design flow that takes chiplets' netlist into consideration. 
However, this analysis would be hard to perform in the chiplet marketplace since the vendors may be reluctant to share their intellectual properties. 
Therefore, the vendors and customers need to find standard models to describe the electrical properties of chiplets and enable fast simulation without revealing any detailed circuit designs as discussed in standards~\cite{Mastroianni2021standard}.

\subsection{Security Issues: Threat Models and Protections} 
The security of chiplet systems also poses significant challenges for both chiplet vendors and designers. 
Here, we focus on the security issues from the product designers' perspective. 
Combining different chiplets from various vendors, chiplet systems are more vulnerable to hardware security threats.
On the one hand, chiplet systems expose more vulnerabilities for attacks~\cite{hu2020overviewofHWsecurity}.
The interaction among chiplets may lead to more system side-effects.
the design flaws or malicious circuits not only appear in the chiplet but also can be inserted in the integration stage or even in an active interposer~\cite{sami2023enabling_security_in_HI}.
On the other hand, the complicated computing architecture and mix-trust environment will bring more difficulties in developing and deploying the protection methods on the chiplet systems.

\subsubsection{Potential threats}

From the product designers' perspective, side-channel attacks, fault-injection attacks, and hardware trojans are common threats, and 
these threats can also happen in chiplet systems.

\noindent\textbf{Side-channel attack} is the attack that utilizes the unintentional leakage information of the computing systems, which includes memory, timing, power, electromagnetic, etc. 
For example, Zombie Load~\cite{schwarz2019zombieload} successfully extracts sealing keys from the Intel SGX~\cite{mckeen2016intelSGX} enclave by using leaking memory data of hyperthreading. 
Recently, some attacks have also been proposed on accelerator systems like GPU systems~\cite{luo2019side-channel_on_GPU} or in-memory computing (IMC) systems~\cite{wang2023side-channel_on_IMC}. 
In chiplet systems, the interconnect channels may become new victims of side-channel attacks. For example, the contention in the interconnect could cause key leakage in cryptographic systems~\cite{dai2022don}.


\noindent\textbf{Fault-injection attacks} use malicious circuits to trigger faults in aspects like voltage~\cite{krautter2018fpgahammer}, temperature~\cite{alam2019ram}, electromagnetic\cite{elmohr2020_EM_fault_injection}, and so on. 
These threats could also exist in the chiplet systems, which enables attackers to manipulate the behavior of the chiplet systems in multiple ways. 

\noindent\textbf{Hardware trojan} is malicious modification or inclusion of additional malicious circuits. 
The hardware trojan could be inserted into the system in each stage of the supply chain including design, implementation, and integration~\cite{sami2023enabling_security_in_HI}. Chiplet systems can be prone to hardware trojans since there are more parts and entities, e.g., untrusted vendors in the supply chain.

\subsubsection{Potential protection methods}

The novel and complicated architecture of chiplet systems also brings challenges in adopting protection methods. We summarize existing efforts that could be adopted in chiplet systems below.

\noindent\textbf{Trusted execution environments (TEEs)} provide an isolated space, namely an enclave, in which programs can be safely executed even in an untrusted CPU. 
Initially, TEEs like Intel SGX~\cite{mckeen2016intelSGX} only protect the basic CPU architecture, while recently more methods have been proposed for heterogeneous architecture with accelerators like GPU~\cite{volos2018graviton_gpu_tee} and FPGA~\cite{zhao2022shef}.
The rise of chiplet systems calls for more advanced methods: a TEE may need to provide a safe environment across multiple chiplets with different types. 
We need a generalized TEE design framework that can be applied to different chiplets, and different D2D interfaces, to minimize the design efforts.

The novel architecture of chiplet systems also brings new methods and opportunities for protection. 
\cite{nabeel20202.5D-ROT} proposed a new concept of ``Root of Trust", which guarantees security by utilizing an active interposer. \cite{safari2023hybrid_obfuscation} 
improves the obfuscation method, which adopts a hybrid split manufacturing methodology and uses different vendors to manufacture the obfuscated circuits. 
Therefore, in chiplet systems, the chip design can be further spilled and obfuscated, thus providing more security. \cite{sami2023enabling_security_in_HI} proposes to integrate additional trusted chiplets such as chiplet-based hardware security modules (CHSM) and chiplet-based security IPs (CSIP). 
The CHSM is an FPGA chiplet from trusted vendors and features many sensors to detect probing or fault injection attacks. 
The reconfigurability of CHSM allows remote security policy updates in response to zero-day attacks. 
CSIP can prevent attackers that have physical access to the system by establishing encrypted communication between chiplets. 
To achieve security, CSIP must be provided by trusted vendors and has cryptographic modules, physically unclonable functions (PUFs), etc.
\section{Software Design Challenges
}

Programming software on chiplet systems is challenging.
In existing host-device systems, CPUs can offload tasks to accelerators such as FPGAs and GPUs and runtime is needed to 
%
manage the data movement and execution. 
To enable the host orchestration, vendor-dependent runtimes are required in the host machine. 
In a chiplet system, the application will probably be executed based on a number of different runtimes. 
If a runtime does not cooperate with others and assumes full ownership of the host system, conflicts may occur which might lead to erroneous behaviors. 
Besides, different accelerators usually need different development environments and design kits.
This means the developers have to write code that runs on each chiplet separately. 
This not only introduces extra design efforts but also reduces the portability of the application. 

A unified programming infrastructure seems a promising solution. 
Though many unified programming methods have been proposed, it is still an open question that which infrastructure fits the chiplet systems best. 
Here we summarize some existing explorations.

In 2014,  SYCL~\cite{SYCL_web} was proposed to achieve heterogeneous device programming for applications. 
As an open industry standard, SYCL utilizes the C++ programming model.
Several implementations of SYCL have been introduced, including Intel oneAPI~\cite{wisniewski2021holistic}. 
oneAPI is designed as a unified development environment for different kinds of accelerators from different vendors. 
Based on the Level Zero API, which acts as the lowest-level interface, oneAPI provides a whole software stack including system software, developer interface, etc.

Besides SYCL, MLIR~\cite{lattner2021mlir} is also a desired choice. 
As a compiler infrastructure, MLIR is adopted in different projects like TensorFlow Graphs and Fortran IR, as well as various domain-specific compilers. 
Some implementations of MLIR on heterogeneous hardware systems are also been made recently. 
ScaleHLS~\cite{ye2021scalehls} uses MLIR in the FPGA high-level synthesis (HLS) to suit the intrinsic hierarchies of HLS design and enables larger design space and more optimization chances. 
HeteroCL~\cite{lai2019heterocl} proposed a programming infrastructure, decoupling algorithm specifications with hardware customization, which gives developers an efficient way to explore larger design space and higher performance.

To fully unleash the performance of chiplet systems, there are many more aspects including higher-level software tools, including how to map and partition the workload onto heterogeneous chiplets, how to do mapping and architecture co-design for chiplet systems, etc.
H2H~\cite{zhang2022h2h} proposes a communication-aware mapping algorithm to map heterogeneous models to heterogeneous systems. 
CHARM~\cite{fpga23charm} proposes a software mapping framework to map heterogeneous kernels within end-to-end deep learning applications and heterogeneous components within an SoC. 
Such discussions should be considered in chiplet system scenarios and reevaluated. 

On top of that, how to perform efficient design space explorations (DSE) in chiplet systems is also much needed.
For heterogeneous systems, the techniques needed for fine-grained control like pipeline or parallelism require a deep understanding of the architecture from the software developers. 
There is a need for tools that can: (1) describe different optimization techniques on different chiplet architectures; (2) automatically find the optimal configurations on these heterogeneous chiplet systems. 
For example, for FPGA accelerators, AutoDSE~\cite{sohrabizadeh2022autodse} develop a DSE framework, aiming to solve the optimization bottleneck, usually application-specific, to enable common software developers to produce high-quality FPGA programs. 
Similar DSE tools are needed in more complicated systems, e.g., the chiplet system.

\section{Conclusion}
With the boom in AI models and fast-increasing demands on performance and energy efficiency, heterogeneous integration based on chiplet becomes a promising way to keep scaling while minimizing the total cost. 
This paper summarizes the current challenges 
and 
calls for innovations and collaborative efforts to address these challenges.

\section*{Acknowledgement}
We acknowledge the support from the University of Pittsburgh New Faculty Start-up Grant and 
National Science Foundation awards \#2213701, \#2217003, \#2324864.


\bibliographystyle{IEEEtran}
\balance
\bibliography{reference}


\end{document}